\documentclass[ste,twoside]{stefano}

\usepackage{amsmath}
\usepackage{amssymb}
\usepackage{graphics}
\usepackage{rotating}
\usepackage{cite}
\usepackage{color}

\def\makeheadbox{{%
\hbox to0pt{\vbox{\baselineskip=10dd\hrule\hbox
to\hsize{\vrule\kern3pt\vbox{\kern3pt \hbox{  {\sc J. Math. Phys.} {\bf
47}, 102104-9  (2006) } \hbox{ {\sc
{\color{blue}{dma}}[{\color{black}{imecc}}]{\color{red}{UniCamp}} }
\hspace*{10.4cm} {\color{blue}{$\boldsymbol{\Sigma \delta \Lambda}$}} }
\kern3pt}\hfil\kern3pt\vrule}\hrule}%
\hss}}}

\def\0{\mbox{\tiny $0$}}
\def\1{\mbox{\tiny $1$}}
\def\2{\mbox{\tiny $2$}}
\def\3{\mbox{\tiny $3$}}
\def\4{\mbox{\tiny $4$}}
\def\5{\mbox{\tiny $5$}}
\def\6{\mbox{\tiny $6$}}
\def\7{\mbox{\tiny $7$}}
\def\8{\mbox{\tiny $8$}}
\def\9{\mbox{\tiny $9$}}
\def\I{\mbox{\tiny $I$}}
\def\II{\mbox{\tiny $II$}}
\def\mi{\mbox{\tiny $-$}}
\def\pl{\mbox{\tiny $+$}}
\def\plmi{\mbox{\tiny $\pm$}}
\def\M{\mbox{\tiny max}}
\def\min{\mbox{\tiny min}}
\def\i{\mbox{\tiny inc}}
\def\r{\mbox{\tiny ref}}
\def\t{\mbox{\tiny tra}}
\def\ic{\mbox{\tiny $(i)$}}
\def\jkh{\mbox{\tiny $(j,k)$}}
\begin{document}
%

\title{\Large QUATERNIONIC DIFFUSION BY A POTENTIAL STEP}

\author{
Stefano De Leo\inst{1}
\and Gisele C. Ducati\inst{2}
}

\institute{
Department of Applied Mathematics, University of Campinas\\
PO Box 6065, SP 13083-970, Campinas, Brazil\\
{\em deleo@ime.unicamp.br}
 \and
Department of Mathematics, University of Parana\\
PO Box 19081, PR 81531-970, Curitiba, Brazil\\
{\em ducati@mat.ufpr.br}
}


\date{Submitted: {\em August, 2006}. Revised: {\em September, 2006}.  }

\abstract{In looking for {\em qualitative} differences between
quaternionic and complex formulations of quantum physical
theories, we provide a detailed discussion of the behavior of a
wave packet in presence of a quaternionic time-independent
potential step. In this paper, we restrict our attention to
diffusion phenomena. For the group velocity of the wave packet
moving in the potential region and for the reflection and
transmission times, the study shows a striking difference between
the complex and quaternionic formulations which could be matter of
further theoretical discussions  and could represent the starting
point for a possible experimental investigation.}


\PACS{ {03.65.Fd} \and {03.65.Nk}{}}












\titlerunning{\sc quaternionic diffusion by a potential step}

\maketitle


\section*{I. INTRODUCTION}
Despite much research on quaternionic quantum mechanics, reviewed
in its mathematical and physical aspects in the excellent book of
Adler\cite{ADL}, there have been few breakthroughs on the most
natural question about the effect that quaternionic potentials
play in the dynamics of elementary
particles\cite{DAV89,DAV92,DEDUC02,DEDUC05} and, as consequence of
it, about the possibility to look for an experimental
proposal\cite{PER79,KAI84,KLE88}. In this paper, by using the new
mathematical tools developed in the analytic resolution of
eigenvalue problems\cite{DESC00,DESC02} and differential
equations\cite{DEDUC01,DEDUC03,DEDUC04,DEDUC06}, we analyze in
detail the diffusion of a wave packet by a quaternionic potential
step.

The fundamental (mathematical) point in our discussion is the use of {\em
real} quaternionic inner products and wave functions. The use of a division
algebra is needed to guarantee that amplitudes of probability satisfy the
requirement that, in the absence of quantum interference effects, the
probability amplitude superposition reduces to the probability
superposition. The associative law of multiplication (which for example
fails for octonions) is needed to satisfy the completeness formula and to
guarantee the invariance of the inner product for anti-self-adjoint
operators\cite{ADL}. This does not mean that we cannot formulate a
consistent quantum theory based on the use of complexified quaternions,
Clifford algebras or octonionions. The requirement of an associative
division algebra only applies to inner products. For example, in
litterature we find interesting formulation of quantum theory based on the
use of wave functions defined in the complexified quaternionic
algebra\cite{CQ1,CQ2,CQ3,CQ4,CQ5a,CQ5b,CQ6}, in the space-time
algebra\cite{ST1,ST2,ST3,ST4,ST5} and in the octonionic
algebra\cite{OC1,OC2} {\em but} with inner products projected over the
complex field. Nevertheless, from our point of view the choice of
quaternionic inner products seems to be the best choice in investigating
deviations from standard quantum theories\cite{SOU1,SOU2,ADLq1,ADLq2}.

For the convenience of the reader and  to facilitate access to the
individual topics, this work is rendered as self-contained as possible. In
Section II, we set up notation and terminology and proceed with the study
of diffusion by quaternionic potentials. This section contains the
(analytic) plane wave solution of the quaternionic Schr\"odinger equation
in the presence of a potential step.  This  represents a fundamental
mathematical tool in the discussion of the quaternionic stationary phase
method (see Section III). We will touch only a few aspects of the theory of
quaternionic integral transforms and restrict our attention to the
diffusion of quaternionic wave packets with a peaked convolution function.
The advantage of using the stationary phase method lies in the fact that,
in the presence of a potential step, the motion of the wave packet can be
correctly estimated by analyzing the phase derivative calculated at the
maximum of the convolution function\cite{MAT,MER,COHEN}. For a different
shape of potentials, see for example the barrier, the stationary phase
method, depending on the width of the potential and on the group velocity
of the incoming particle, could break down. There is a rich number of
articles leading with this problem in standard quantum
mechanics\cite{HAR,MUL1,HAR2,MUL2,OREC,DIR1,DIR2}.

The results of this paper (conclusion and outlooks are drawn in
Section IV) shed some new light on the properties of quaternionic
potentials. In particular, it is explicitly shown how the presence
of a quaternionic perturbation modifies the momentum of the
non-relativistic incoming particle and its reflection
(transmission) time. The study presented in this article
represents a starting point in view of a complete understanding of
the behavior of wave packets impinging on quaternionic potentials.
A detailed analysis of this topic could be fundamental  in looking
for experiments in which deviations from the complex quantum
theory could be really seen. It is worth pointing out that the
question of finding the best experimental proposal to prove the
existence of quaternionic potentials is, at present, far from
being solved and this paper aims to contribute to this debate.

\section*{II. REFLECTION AND TRANSMISSION COEFFICIENTS}

The quaternionic Schr\"odinger equation in the presence of a
constant  potential is given by
\begin{equation}
\label{sceq} i\,\frac{\hbar}{2m}\,\Psi_{xx}(x,t)-\frac{i\,V_{\1}
+j\,V_{\2}+ k\, V_{\3}}{\hbar} \,\Psi(x,t) =  \Psi_t(x,t)\, \,,
\end{equation}
where $i\,V_{\1} +j\,V_{\2}+ k\, V_{\3}$ represents the
quaternionic generalization of the anti-hermitian complex
potential $i\,V_{\1}$. For a complete discussion see
ref.\cite{ADL}. This partial differential equation, by the
substitution
\begin{equation}
\label{sta} \Psi(x,t) = \Phi(x)\, \exp[-\,i\,E\,t/\hbar]\,\, ,
\end{equation}
can be reduced to the following ordinary second order
differential equation with constant quaternionic coefficients,
\begin{equation}
i\,\frac{\hbar^{\2}}{2m}\,\Phi''(x)-\left(\,i\,V_{\1} +j\,V_{\2}+
k\, V_{\3}\, \right)\,\Phi(x) = - \, \Phi(x)\,E\, i\,\,.
\end{equation}
The solution of the Schr\"odinger equation in the presence of
constant quaternionic  potential has been matter of study in the
recent years\cite{DEDUC02,DEDUC05,DEDUC06}. New mathematical
techniques, essentially based on the right eigenvalue problem for
quaternionic operators\cite{DESC00,DEDUC01}, allow to obtain the
solution without the need to translate the quaternionic problem in
its complex counterpart\cite{DAV89,DAV92}. In particular, in the
presence of a potential step and for the diffusion case
\[ E >
\sqrt{V_{\1}^{^{\2}}+V_{\2}^{^{\2}}+V_{\3}^{^{\2}}}\,\, ,\] the
quaternionic plane wave solutions (for a detailed derivation see
ref.\cite{DEDUC06}) are:
\begin{equation}
\label{wf}
\begin{array}{llcl}
\mbox{\sc [I] free region ($x<0$) :} \hspace*{.5cm}  &
\Phi_{\I}(x) & = & \exp[\,i\,\epsilon\,x\,] + r \, \exp[\,-\,i\,
\epsilon \,x\,] +
j\, \tilde{r}\, \exp[\,\epsilon\, x\,]\,\, , \\ \\
\mbox{\sc [II] potential region ($x>0$) :} \hspace*{.5cm}  &
\Phi_{\II}(x) & = & (1+j\, w)\, t\,\exp[\,i\,\rho_{\mi}\, x\,] +
(\bar{w}+j)\, \tilde{t}\,\exp[\,-\,\rho_{\pl} \,x\,]\,\, ,
\end{array}
\end{equation}
where
\[
\epsilon   =  \sqrt{ \,\displaystyle{ \frac{2m}{\hbar^{^{\2}}}}\,
E}\,\,\,,\,\,\,\,\,\rho_{\plmi} = \sqrt{ \,\displaystyle{
\frac{2m}{\hbar^{^{\2}}}}\,\left(
\sqrt{E^{^{\2}}-V_{\2}^{^{\2}}-V_{\3}^{^{\2}}} \pm
V_{\1}\right)}\,\,\, ,\,\,\,\,\, w = -\, i \,
\displaystyle{\frac{V_{\2}-iV_{\3}}{E +
\sqrt{E^{^{\2}}-V_{\2}^{^{\2}}-V_{\3}^{^{\2}}}}}\,\,,
\]
and
\[
\begin{array}{lclclclc}
t & = & \displaystyle{\frac{2\epsilon}{\epsilon +
\rho_{\mi}}}\,\left[\,1-\, |w|^{\2} \, \frac{\epsilon -
i\rho_{\mi}}{\epsilon+\rho_{\mi}}\, \frac{\epsilon +
i\rho_{\pl}}{\epsilon+\rho_{\pl}}\,\right]^{\mi \, \1}& \,\, ,
\,\,\,\,\,\,\,\,& \tilde{t} & = &\displaystyle{\frac{i\rho_{\mi} -
\epsilon}{\epsilon+\rho_{\pl}}}\,\,\,w\,t & \, \,\,,\\
r & = & \displaystyle{\frac{\epsilon -
\rho_{\mi}}{2\epsilon}}\,\left[\,1-\, |w|^{\2}\, \frac{\epsilon -
i\rho_{\mi}}{\epsilon-\rho_{\mi}}\, \frac{\epsilon -
i\rho_{\pl}}{\epsilon+\rho_{\pl}}\,\right]\, t&\,\, ,
\,\,\,\,\,\,\,\,& \tilde{r} & = &\displaystyle{\frac{i\rho_{\mi}+
\rho_{\pl}  }{\epsilon+\rho_{\pl}}}\,\,\,w\,t & \,\,\,.
\end{array}
\]
From the current conservation
\begin{equation}
 \left[\,\overline{\Psi}(x,t)\Psi(x,t)\,\right]_t =
\frac{\hbar}{2\,m}\,\left[\, \overline{\Psi}(x,t) \,i\,
\Psi_{xx}(x,t) + \mbox{h.c}\,\right]\,\, ,
\end{equation}
by recalling that we are considering stationary solutions  of the
Schr\"odinger equation, we obtain
\[ \overline{\Phi}(x) \,i\,
\Phi''(x) + \, \mbox{h.c.}= 0\,\,.\] This implies that the current
density,
\begin{equation}
\label{cur}
 \mbox{J}(x)= \overline{\Phi}(x) \,i\, \Phi'(x) +\,
\mbox{h.c}\,\,,
\end{equation}
is a quantity independent of $x$. Due to the continuity of the
wave function and its derivative, the current density has to
satisfy the following constraint
\begin{equation}
\label{con}
\mbox{J}_{\I}(0)=\mbox{J}_{\II}(0)\,\, .
\end{equation}
By using the explicit form of the plane wave solutions given in
Eqs.(\ref{wf}), and the condition (\ref{con}), a straightforward
calculation conduces to
\begin{equation}
R + T = 1
\end{equation}
where
\[ R=|r|^{\2}\,\,\,\,\,\mbox{and}\,\,\,\,\,
T=\frac{\rho_{\mi}}{\epsilon}\,\,\left( 1 -
|w|^{\2}\right)\,|t|^{\2}\,\,.\] Similarly to the predictions of
complex quantum mechanics, the incident particle has a non-zero
probability of turning back. Nevertheless, we know that in
standard quantum mechanics no phase  is created by such
reflection\cite{COHEN}. The situation drastically changes in the
presence of a quaternionic perturbation. We shall come back to
this point in Section III.

\subsection*{II.A REFLECTION AND TRANSMISSION PHASES}
From the stationary wave functions given in Eqs.(\ref{wf}),  we
shall construct, by linear superposition, wave packets and we
shall study their time evolution (see Section III). In this
spirit, it is convenient to rewrite the reflection and
transmission coefficients in terms of their modulus and phases. By
simple algebraic manipulations, we find
\begin{eqnarray}
r & = & \sqrt{\frac{ [(\epsilon - \rho_{\mi})(\epsilon +
\rho_{\pl}) - |w|^{\2} \,(\epsilon^{\2} -
\rho_{\mi}\rho_{\pl})]^{^{\2}} + |w|^{\4} \epsilon^{\2} (
\rho_{\mi}+\rho_{\pl})^{^{\2}} }{ [(\epsilon +
\rho_{\mi})(\epsilon + \rho_{\pl}) - |w|^{\2}\,(\epsilon^{\2} +
\rho_{\mi}\rho_{\pl})]^{^{\2}} + |w|^{\4}\epsilon^{\2} (
\rho_{\mi}-\rho_{\pl})^{\2} }}\, \,\,
\exp[\,i\,\theta_r\,]\,\,, \nonumber \\
 t & = &
\frac{2\,\epsilon\, (\,\epsilon + \rho_{\pl}\,)}{ \sqrt{[(\epsilon
+ \rho_{\mi})(\epsilon + \rho_{\pl}) - |w|^{\2}\,(\epsilon^{\2} +
\rho_{\mi}\rho_{\pl})]^{^{\2}} + |w|^{\4}\epsilon^{\2} (
\rho_{\mi}-\rho_{\pl})^{\2} }}\, \,\, \exp[\,i\,\theta_t\,]\,\,,
\end{eqnarray}
where
\begin{eqnarray}
\theta_r & = &  \arctan \left[ \frac{ \epsilon\, (
\rho_{\pl}+\rho_{\mi})\,|w|^{\2} }{(\epsilon -
\rho_{\mi})(\epsilon + \rho_{\pl}) - |w|^{\2} \,(\epsilon^{\2} -
\rho_{\mi}\rho_{\pl})}\right] + \,\theta_t\,\,, \nonumber
\\
\theta_t& = &\arctan \left[ \frac{ \epsilon\, (
\rho_{\pl}-\rho_{\mi})\,|w|^{\2} }{(\epsilon +
\rho_{\mi})(\epsilon + \rho_{\pl}) - |w|^{\2} \,(\epsilon^{\2} +
\rho_{\mi}\rho_{\pl})}\right]\,\,.
\end{eqnarray}
The important point to be noted here is the dependence on the
energy, $E$, and the complex imaginary part of the potential,
$V_{\1}$, as expected from the standard quantum case, and the {\em
new} dependence on the modulus of  the pure quaternionic part of
the  potential, $|V_{\2}+i\, V_{\3}|$. This last result means that
once fixed the modulus of the quaternionic perturbation any
rotation in the plane $(V_{\2}\,,V_{\3})$ does not modify the
reflection and transmission coefficients. The quaternionic
rotation invariance is due to the choice of $i$ as the imaginary
unit  in the anti-hermitian momentum operator,
$i\hbar\,\partial_{xx}/2m$, which appears in the quaternionic
Scrh\"odinger equation (\ref{sceq}).

\subsection*{II.B THE COMPLEX LIMIT}
The standard (complex) quantum results can be obtained by taking a
simple limit case, i.e. $V_{\2,\3}\to 0$. In fact, by observing
that
\[
V_{\2,\3}\to 0\,\,\,\Rightarrow\,\,\,\left\{ \begin{array}{lcl} w
& \to & 0 \,\, ,\\
\rho_{\mi} & \to & \sigma= \sqrt{ \,\displaystyle{
\frac{2m}{\hbar^{^{\2}}}}\, \left(E-V_{\1}\right)} \,\,,
\end{array} \right.
\]
we find
\[ R=|r|^{\2}\,\,\,\,\,\mbox{and}\,\,\,\,\,
T=\frac{\,\sigma}{\epsilon}\,|t|^{\2}\,\, ,\] where
\begin{equation}
r =  \frac{\epsilon -\sigma}{\epsilon + \sigma}
\,\,\,\,\,\mbox{and}\,\,\,\,\,
 t= \frac{ 2\,\epsilon}{\epsilon + \sigma}\,\,.
\end{equation}
As expected, the reflection and transmission coefficients ($r$ and
$t$) are real $(\theta_{r}=  \theta_{t}=0)$ and this implies that
there is no phase created by reflection or
transmission\cite{COHEN}.

\subsection*{II.C THE PURE QUATERNIONIC LIMIT}
It is interesting to consider a second limit, i.e. $V_{\1}\to 0$.
This represents the case of a {\em pure} quaternionic potential.
Noting that
\[
V_{\1}\to 0\,\,\,\Rightarrow\,\,\,\left\{ \begin{array}{lcl}
\rho_{\plmi}  & \to &  \rho= \sqrt{ \,\displaystyle{
\frac{2m}{\hbar^{^{\2}}}}\,
\sqrt{E^{^{\2}}-V_{\2}^{^{\2}}-V_{\3}^{^{\2}}}} \,\,,\\
 |w|^{\2} & \to & \displaystyle{
\frac{\epsilon^{\2} - \rho^{\2}}{\epsilon^{\2} + \rho^{\2}}}\,\, ,
\end{array} \right.
\]
we obtain
\[ R=|r|^{\2}\,\,\,\,\,\mbox{and}\,\,\,\,\,
T=\frac{2\,\rho^{\3}}{\epsilon\,(\epsilon^{\2}+\rho^{\2})}\,|t|^{\2}\,\,,
\]
where
\begin{equation}
r =  \frac{\,\,\epsilon - \rho}{ \sqrt{\epsilon^{\2} +
\rho^{\2}}}\,\exp\left[\,i\,\arctan
\left(\,\frac{\epsilon}{\rho}\,\right)\,\right]
\,\,\,\,\,\mbox{and}\,\,\,\,\,
 t  = \frac{ \epsilon}{\rho}\,\,.
\end{equation}
In this limit, the symmetry between reflection and transmission
times is broken down. For a pure quaternionic potential step, we
find an instantaneous transmission but {\em not} an instantaneous
reflection (we shall discuss in detail this point in Section III).

\section*{III. STATIONARY PHASE METHOD}

Until now, we have been concerned only with plane waves. In this
Section, we are going to study the time evolution of quaternionic
wave packets and deducing from them several important properties.
The principle of superposition guarantees that every {\em real}
linear combination of the plane waves
$\Phi_{\I}(x)\exp[-iEt/\hbar]$ and $\Phi_{\II}(x)\exp[-iEt/\hbar]$
will satisfy the Schr\"odinger equation in the presence of a
quaternionic potential step.

Let $g(\epsilon)$ be a real convolution function with a maximum in
$\epsilon_{\0}$. In the free region ($x < 0$), the superposition
can be written as follows
\begin{equation}
\label{omega1}
 \Omega_{\I}(x,t) =
\int_{\epsilon_{\min}}^{\,\infty} \hspace*{-.2cm}\mbox{d}\epsilon
\,\,g(\epsilon)\, \left\{\,\exp[\,i\,\epsilon\,x\,] + r \,
\exp[\,-\,i\,\epsilon \,x\,] + j\, \tilde{r}\, \exp[\,\epsilon\,
x\,]\,\right\} \,\exp[\,-\,i\, \epsilon^{\2} \, \hbar
\,t/\,2m\,]\,\,,
\end{equation}
where
\[\epsilon_{\min}=\sqrt{\frac{2m}{\hbar^{^{\2}}}\,
\sqrt{V_{\1}^{^{\2}}+V_{\2}^{^{\2}}+V_{\3}^{^{\2}}}}\,\,.\] The
first term in Eq.(\ref{omega1}) represents the incident wave, the
second term the reflected wave and the third term an evanescent
wave. The phases for the incoming and reflected waves are
\begin{eqnarray}
\theta_{\i}[\epsilon;x,t] & = & \epsilon \,x  - \epsilon^{\2}
\frac{\hbar\,t}{2m}\,\,,\nonumber \\
\theta_{\r}[\epsilon;x,t]  & = & - \,\epsilon \,x - \epsilon^{\2}
\frac{\hbar\,t}{2m} + \, \theta_r\,\,.
\end{eqnarray}
The stationary phase condition (the derivative with respect to
$\epsilon$ of the argument calculated in $\epsilon_{\0}$ equal to
zero) enables us to calculate the position of the maximum of the
incident and reflected wave packets:
\begin{eqnarray}
\label{xir}
x^{\M}_{\i}(t) & = & \frac{\hbar \,\epsilon_{\0}}{m}\,t\,\, , \nonumber\\
 x^{\M}_{\r}(t) & = & -\frac{\hbar \,\epsilon_{\0}}{m}\,t+
\left[\,\frac{\,\,\mbox{d} \theta_r}{\mbox{d}
\epsilon}\,\right]_{\0}\,\, .
\end{eqnarray}
The maximum of the incident wave packet arrives at the step
discontinuity at time $t=0$ (as it happens in the complex case).
During a certain interval of time, the wave packet is localized in
the region $x\sim 0$. For large times the incident wave packet has
practically disappeared and we only find the reflected wave
packet. It is important to observe that contrary to the
predictions of complex quantum mechanics $(\theta_r=0)$, the
maximum of the reflected wave packets is found at $x=0$ at time
$t=(m/\hbar \epsilon_{\0})\, [\mbox{d} \theta_r/\mbox{d}
\epsilon]_{\0}$. This means that in the presence of a quaternionic
perturbation we do {\em not} have an instantaneous reflection: for
large times the maximum of the reflected wave packet is {\em not}
at $-\hbar \,\epsilon_{\0}\,t/m$ but is shifted with respect to
this value by a quantity equal to $[\mbox{d} \theta_r / \mbox{d}
\epsilon]_{\0}$.

An analogous discussion for the transmitted wave packet ($x > 0$),
\begin{eqnarray}
\label{omega2} \Omega_{\II}(x,t)  & =  &  \int \mbox{d}\epsilon
\,\,g(\epsilon)\,\left\{\, t\,\exp[\,i\,\rho_{\mi}\, x\,] +
\bar{w}\, \tilde{t}\,\exp[\,-\,\rho_{\pl} \,x\,] \,\right\}\,
\exp[\,-\,i\,
\epsilon^{\2} \, \hbar \,t/\,2m\,]  + \nonumber \\
 & & j\, \int \mbox{d}\epsilon
\,\,g(\epsilon)\,\left\{\, w\, t\,\exp[\,i\,\rho_{\mi}\, x\,] +
\tilde{t}\,\exp[\,-\,\rho_{\pl} \,x\,]\,\right\}\, \exp[\,-\,i\,
\epsilon^{\2} \, \hbar \,t/\,2m\,]\,\,,
\end{eqnarray}
where the phases to be considered are
\begin{eqnarray}
\theta_{\t}^{\mbox{\tiny $(1,i)$}}[\epsilon;x,t] & = &
\rho_{\mi}\, x - \epsilon^{\2}
\frac{\hbar\,t}{2m} + \theta_t \nonumber\\
\theta_{\t}^{\mbox{\tiny $(j,k)$}}[\epsilon;x,t] & = &
\rho_{\mi}\, x - \epsilon^{\2} \frac{\hbar\,t}{2m} + \theta_t +
\arctan\left[ \frac{V_{_{\2}}}{V_{_{\3}}}\right]\,\,,
\end{eqnarray}
leads to a similar conclusion for the transmitted time. Contrary
to what happens in the standard (complex) quantum mechanics where
there is an instantaneous transmission, in the presence of a
quaternionic potential step, the maximum of the transmitted wave
packet,
\begin{equation}
\label{xt} x^{\M}_{\t}(t) = \left\{\,\frac{\hbar
\,\epsilon_{\0}}{m}\,t - \left[\,\frac{\,\,\mbox{d}
\theta_t}{\mbox{d} \epsilon}\,\right]_{\0} \,
\right\}\,\mbox{\Large $/$} \,\left[\,\frac{\,\,\,\,\mbox{d}
\rho_{\mi}}{\mbox{d} \epsilon}\,\right]_{\0}\,\,,
\end{equation}
is found at $x=0$ at time $t=(m/\hbar \epsilon_{\0})\, [\mbox{d}
\theta_t/\mbox{d} \epsilon]_{\0}$. At first glance, it could
appear a logical consequence of the result obtained for the
reflection time. Nevertheless, it is important to note that
$\theta_r\neq \theta_t$ and consequently the symmetry between
reflection and transmission times is always broken down. For
example, as it was explicitly shown in the previous section,
instantaneous transmission does not necessarily imply
instantaneous reflection.

In order to simplify the discussion about the results obtained in
our study, let us introduce the following notation
\[
V_{\0}=\sqrt{V_{\1}^{^{\2}} + V_{\2}^{^{\2}}+V_{\3}^{^{\2}}}\,\,,
\]
and rewrite the maximum of the incident, reflected and transmitted
wave packets in terms of $E_{\0}$ (the maximum value of the energy
spectrum of the incoming particles)
\begin{eqnarray}
\label{for}
x^{\M}_{\i}(t) & = & \sqrt{\frac{2E_{\0}}{m}}\,\,t\,\,,\nonumber \\
 x^{\M}_{\r}(t)& = & - \sqrt{\frac{2E_{\0}}{m}}\,\,t +
 \frac{\hbar}{\sqrt{2mV_{\0}}}\,\,\,\left[\,\frac{\,\,\mbox{d}
\theta_r}{\mbox{d} \sqrt{\frac{E}{V_{\0}}}}\,\right]_{\0}\,\,,
\nonumber \\
x^{\M}_{\t}(t) & = &\left\{\, \sqrt{\frac{2E_{\0}}{m}}\,\,t -
\frac{\hbar}{\sqrt{2mV_{\0}}}\,\,\,\left[\,\frac{\,\,\mbox{d}
\theta_t}{\mbox{d} \sqrt{\frac{E}{V_{\0}}}}\,\right]_{\0}
\,\right\}\, \,\mbox{\huge $/$}
\,\left[\,\frac{\,\,\,\,\mbox{d}\rho_{\mi} }{\mbox{d} \epsilon
}\,\right]_{\0}\,\,.
\end{eqnarray}
The incident and reflected wave packets propagate, respectively,
with velocities of $v_{\0}$ and $-v_{\0}$,
\begin{equation}
v_{\0}=\sqrt{2E_{\0}/m}\,\,. \end{equation} This is the standard
result obtained in complex quantum mechanics. For  the transmitted
wave packet the velocity is given by
\begin{equation}
\label{vel} v_{\t}=v_{\0} \,\mbox{\huge
$/$}\,\left[\,\frac{\,\,\,\,\mbox{d}\rho_{\mi} }{\mbox{d}
\epsilon}\,\right]_{\0}\,\,.
\end{equation}
Due to the fact that the quantity $\rho_{\mi}$ has an additional
dependence on $|V_{\2}+iV_{\3}|$ with respect to the standard
dependence on $V_{\1}$, the complex and quaternionic formulations
give different predictions. For example,  of particular interest,
it is the comparison between the group velocity of the transmitted
wave packet for the complex case, $V_{\0}=V_{\1}$,
\begin{equation}
\label{veli} v_{\t}^{\ic}
 =  v_{\0} \,\mbox{\huge
$/$}\, \left(\, \sqrt{\epsilon^{\2} -
\frac{2m}{\hbar^{^{\2}}}\,V_{\0}}\,\right)_{\epsilon}= v_{\0}
\,\sqrt{1-\frac{V_{\0}}{E_{\0}}}\,\, ,
\end{equation}
 and that one for the {\em pure} quaternionic case,
 $V_{\0}=|V_{\2}+iV_{\3}|$,
\begin{equation}
\label{veljk} v_{\t}^{\jkh}
 =  v_{\0} \,\mbox{\huge
$/$}\,\left\{\,\left[\,\epsilon^{\4} -
\left(\frac{2m}{\hbar^{^{\2}}}\,V_{\0}\right)^{\2}\,\right]^{\1/\4}\right\}_{\epsilon}
= v_{\0} \,\left[\,1-\left(
\frac{V_{\0}}{E_{\0}}\right)^{\2}\,\right]^{\3/\4}\,\,.
\end{equation}
A first {\em important} observation is that whereas $v_{\t}^{\ic}$
is greater or smaller than $v_{\0}$ depending on the sign of
$V_{\1}$, $v_{\t}^{\jkh}$ is always smaller than the group
velocity in the fre region. For incident particles with an energy
spectrum peaked in $E_{\0}$, with $E_{\0}\gg V_{\0}$, the group
velocities of the wave packet travelling in the potential region,
(\ref{veli}) and (\ref{veljk}), can be approximated by taking the
first terms in their Taylor expansions,
\[
\begin{array}{lcl}
v_{\t}^{\ic} & \,= \,\, &  v_{\0}\, \left[\displaystyle{1
-\frac{1}{2}\,\frac{V_{\0}}{E_{\0}} - \frac{1}{8}\, \left(
\frac{V_{\0}}{E_{\0}}\right)^{\2}}\,\right] +
\displaystyle{\mathcal{O}\left[\left( \frac{V_{\0}}{E_{\0}}\right)^{\3}\,\right]} \,\,,\\
v_{\t}^{\jkh} & \,= \,\, & v_{\0}\,\left[ \displaystyle{1 -
\frac{3}{4}\, \left( \frac{V_{\0}}{E_{\0}}\right)^{\2}}\,\right] +
\displaystyle{\mathcal{O}\left[\left(
\frac{V_{\0}}{E_{\0}}\right)^{\4}\,\right]} \,\,.
\end{array}
\]
This means that a clear difference  between the complex and the
(pure)  quaternionic case is expected for the group velocity of a
wave packet travelling in a region in which a small pertrbation is
turned on. In this spirit, it is also interesting to compare the
reflection and transmission times,
\begin{eqnarray} \label{times} \frac{\sqrt{2mV_{\0}}}{\hbar}\,\,
x^{\M}_{\r}(0)& = & \left[\,\frac{\,\,\mbox{d} \theta_r}{\mbox{d}
\sqrt{\frac{E}{V_{\0}}}}\,\right]_{\0}\,\,, \nonumber
\\
\frac{\sqrt{2mV_{\0}}}{\hbar}\,\,x^{\M}_{\t}(0) & = & -
\left[\,\frac{\,\,\mbox{d} \theta_t}{\mbox{d}
\sqrt{\frac{E}{V_{\0}}}}\,\right]_{\0} \,\mbox{\huge $/$}\,
\left[\,\frac{\,\,\,\,\mbox{d}\rho_{\mi} }{\mbox{d} \epsilon
}\,\right]_{\0}\,\,.
\end{eqnarray}
Standard quantum mechanics predicts instanteneous reflection and
transmission, i.e.
\[ x^{\ic \, \M}_{\r}(0)=x^{\ic \, \M}_{\t}(0)=0\,\,.\]
For a pure quaternionic potential, the transmission, in analogy to
the complex case, is instantaneous,
\[ x^{\jkh \, \M}_{\t}(0)=0\,\, ,\]
 but
the reflection time is different from zero (breaking down the
instantaneity),
\[
\frac{\sqrt{2mV_{\0}}}{\hbar}\,\, x^{\jkh
\,\M}_{\r}(0)=-\,\left\{\, 2\,\sqrt{\frac{E_{\0}}{V_{\0}}}\,
\left[\,\frac{E_{\0}}{V_{\0}}+ \sqrt{\left(
\frac{E_{\0}}{V_{\0}}\right)^{\2}- 1}\,\right]\,\left[\, \left(
\frac{E_{\0}}{V_{\0}}\right)^{\2} - 1
\,\right]^{\3/\4}\,\right\}^{\mi \1}\,\,.
\]
This predicts, for large times, that the maximum of the reflected
wave packet should be found at the left of the position predicted
by standard quantum mechanics, i.e. $-\hbar \,\epsilon_{\0}\,t/m$.
For $E_{\0}\gg V_{\0}$, the difference between the complex and
(pure) quaternionic case is {\em only} manifest at the third order
in $V_{\0}/E_{\0}$,
\[
\frac{\sqrt{2mV_{\0}}}{\hbar}\,\, x^{\M}_{\r}(0)= - \,
\frac{1}{4}\, \left( \frac{V_{\0}}{E_{\0}}\right)^{\3} +
\mathcal{O}\left[\left( \frac{V_{\0}}{E_{\0}}\right)^{\4}\,\right]
\,\,,
\]
and consequently, for small perturbations, we practically  find an
instantaneous reflection. It is important to note that the shift
in the position of the maximum of the reflected wave packet
becomes important when $E_{\0}$ approaches to $V_{\0}$, this
implies $x^{\jkh \,\M}_{\r}(0)\to \infty$. Nevertheless, for
incident wave packets peaked in $E_{\0}\sim V_{\0}$, a more
careful analysis is needed. In fact, in this limit new effects
have to be considered and these effects cannot be obtained by
simply using the stationary phase method\cite{MUL1,MUL2,OREC}.

\section*{IV. CONCLUSIONS AND OUTLOOKS}
The study presented in this paper, and based on the use of wave
packets,   represents, from our point of view, a first important
attempt to discuss deviations from the standard (complex) quantum
mechanics in the presence of quaternionic potentials. The wave
packet formalism, with respect to the previous analysis,
essentially based on the plane wave solutions, surely gives  a
more "physical" focus. For example, this formalism allows to
explicitly show the effect that quaternionic perturbations play in
the momentum distribution of elementary particles and, in the
particular case of a potential step, to calculate the new
reflection and transmission times due to quaternionic interference
phenomena. To emphasize the main differences beteween the complex
and the quaternionic formulation of quantum mechanics for
diffusion phenomena by a potential step, we have given, in the
previous section, a detailed discussion based on the {\em
analytic} study of the group velocities in the potential region
and of the reflection time for complex and (pure) quaternionic
potentials.

Now, let us return to the discussion for the general case, i.e. a
complex potential in presence of a quaternionic perturbation. In
Fig.1-a, fixed the value of $V_{\0}$ and varying its complex
component $V_{\1}$, we draw
\begin{equation}
\label{rap} \frac{ v_{\t}- v_{\t}^{\ic}}{ v_{\t}^{\ic}}
\end{equation} as a function of $E_{\0}/V_{\0}$. The continuous
line represents the case of a small complex component in the
quaternionic  potential, consequently such a curve approximates
the case of a pure quaternionic potential,
\[
\frac{ v_{\t}^{\jkh}- v_{\t}^{\ic}}{ v_{\t}^{\ic}}=
\left(\,1-\frac{V_{\0}}{E_{\0}}\,\right)^{\1/\4}
\left(\,1+\frac{V_{\0}}{E_{\0}}\,\right)^{\3/\4} -
1\,\,\,\,\,\begin{array}{c} \longrightarrow\\ E_{\0}\gg V_{\0}
\end{array}\,\,\,\,\,\displaystyle{\frac{1}{2}\,
\frac{V_{\0}}{E_{\0}}- \frac{3}{8}\, \left(
\frac{V_{\0}}{E_{\0}}\right)^{\2}} +
\displaystyle{\mathcal{O}\left[\left(
\frac{V_{\0}}{E_{\0}}\right)^{\3}\,\right]}\,\,.
\]
It is also interesting to observe that the maximum of $
(v_{\t}^{\jkh}- v_{\t}^{\ic})/ v_{\t}^{\ic}$ is found at
$E_{\0}=2\,V_{\0}$.

The plots in Fig.1-b and Fig.1-c respectively show the behavior of the
transmission and reflection times as a function of $E_{\0}/V_{0}$. Let us
list some results coming out from our analysis. The quaternionic
interefernce phenomena at the step discontinuity produce an new interesting
effect in the reflected and transmitted wave packets: the maxima of such
packets are found at $x=0$ before that the incident wave packet reaches the
potential step discontinuity. The symmetry between reflection and
transmission time is broken down (see the amplification in Fig.1-b and
Fig.1-c).

Evidently, all the physical consequences of our analysis
regardless of whether we use a complex or a quaternionic potential
in the Schr\"odinger equation deserve further investigation.
Nevertheless, we think that the discussion presented in this paper
and based on the use of the wave packet formalism represents the
starting point for further theoretical studies and a fundamental
tool in looking for possible experimental deviations from standard
(complex) quantum mechanics.

\newpage

\begin{figure}[hbp]
\hspace*{-2.5cm}
\includegraphics[width=19cm, height=22cm, angle=0]{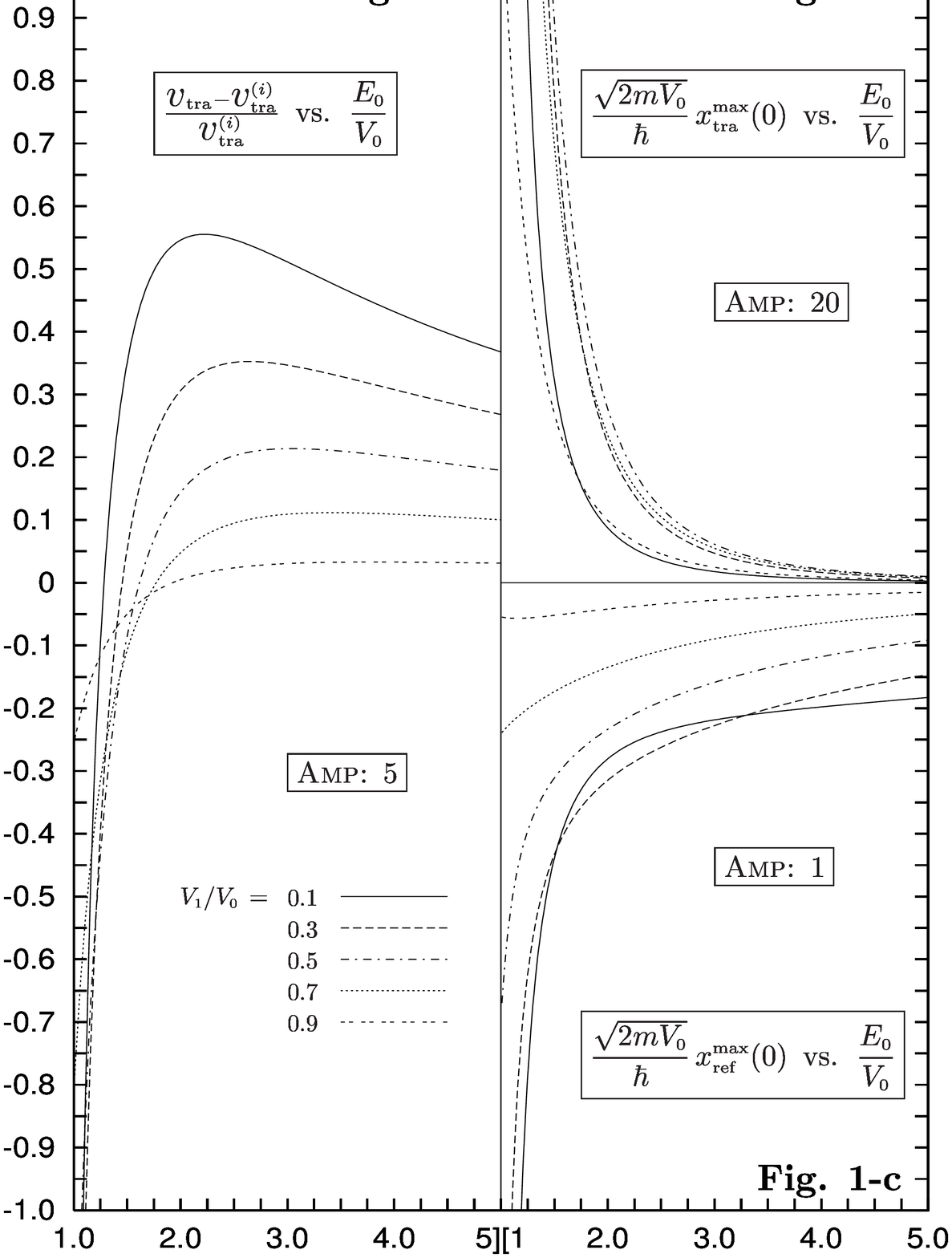}
\caption{Fixed the value of $V_{\0}=|iV_{\1}+jV_{\2}+kV_{\3}|$ and
varying its complex component $V_{\1}$, the group velocity of the
transmitted wave packet, Fig.1-a, and the transmission/reflection
times, Fig.1-b and Fig.1-c, are plotted as a function of
$E_{\0}/V_{\0}$, where $E_{\0}$ is the maximum of the energy
spectrum of the incident wave packet. The analysis is done for
diffusion phenomena ($E_{\0}>V_{\0}$).}
\end{figure}

\end{document}